\documentstyle{article}

\newskip\humongous \humongous=0pt plus 1000pt minus 1000pt

\newif\ifdtup

\def\oldreffmt#1{\rlap{[#1]} \hbox to 2\parindent{}}

\def\figfmt#1{\rlap{Figure {#1}} \hbox to 1in{}}

\def\beq{\begin{equation}}
\def\eeq{\end{equation}}
\def\eq{\beq\eeq}
\relax

\font\tenrm=cmr10

\font\twelvebf=cmbx10 scaled\magstep 2
\font\elevenbf=cmbx10 scaled\magstep 1
\font\elevenrm=cmr10 scaled\magstep 1
\font\elevenit=cmti10 scaled\magstep 1

\renewenvironment{thebibliography}[1]
 { \elevenrm
   \begin{list}{\arabic{enumi}.}
    {\usecounter{enumi} \setlength{\parsep}{0pt}
     \setlength{\itemsep}{3pt} \settowidth{\labelwidth}{#1.}
     \sloppy
    }}{\end{list}}

\newcommand{\npb}[3]{{\elevenit Nucl. Phys.} {\elevenbf #1}, #2 #3}
\newcommand{\plb}[3]{{\elevenit Phys. Lett.} {\elevenbf #1}, #2 #3}
\newcommand{\prd}[3]{{\elevenit Phys. Rev. } {\elevenbf #1}, #2 #3}
\newcommand{\prl}[3]{{\elevenit Phys. Rev. Lett.} {\elevenbf #1}, #2 #3}
\newcommand{\zpc}[3]{{\elevenit Z. Phys.} {\elevenbf #1}, #2 #3}
\newcommand{\prep}[3]{{\elevenit Phys. Rep.} {\elevenbf #1}, #2 #3}

\def\gtrsim{\ \rlap{\raise 2pt \hbox{$>$}}{\lower 2pt \hbox{$\sim$}}\ }
\def\lesssim{\ \rlap{\raise 2pt \hbox{$<$}}{\lower 2pt \hbox{$\sim$}}\ }

\def\hbup{\hfill\break}

\def\ea{{\elevenit et.al.}}
\def\ib{{\elevenit ibid.\ }}
\font\scal=cmsy5

\def\N{{\hbox{\scal N}}}
\def\K{{\hbox{\scal K}}}

\font\mbf=cmmib10  

\def\bfmu{{\hbox{\mbf\char'026}}}

\def\bfe{{\hbox{\mbf\char'145}}}

\def\beq{\begin{equation}}
\def\eeq{\end{equation}}
\def\beqa{\begin{eqnarray}}
\def\eeqa{\end{eqnarray}}

\def\eq{\label}

\begin{document}
\rightline{UM-TH 92--30}\par\noindent
\rightline{hep-ph@xxx/9211246}\par\noindent
\vglue 1.5cm
\begin{center}{{\twelvebf NEW FLAVOR CHANGING INTERACTIONS\\
               \vglue 10pt
               IN EXTENDED GAUGE MODELS$\phantom{|}^\dagger$\\}
\vglue 2.0cm
{\elevenrm ENRICO NARDI \\}
\baselineskip=14pt
\vglue 0.3cm
{\elevenit Randall Laboratory of Physics, University of Michigan\\}
\baselineskip=13pt
{\elevenit Ann Arbor, MI 48109-1120, U.S.A.\\}
\vglue 2.0cm
{\elevenrm ABSTRACT}}
\end{center}
\vglue 0.6cm
{\rightskip=3pc
 \leftskip=3pc
 \elevenrm\baselineskip=14pt
 \noindent
A new class of flavor changing (FC) neutral interactions can arise in
models based on extended gauge groups (rank $>$4) when new charged
fermions are present together with new neutral gauge bosons. In some
cases the FC couplings to a new $Z_1$ are expected to be
sizeable, implying that the $Z_1$ mass must be large enough as to
explain the observed suppression of FC transitions. Concentrating on
E$_6$ models and assuming for the FC parameters a theoretically natural
range of values, I show that in most cases the presence of a
$Z_1$ much lighter than 1 TeV is unlikely.
\vglue 0.6cm}
\noindent
PACS number(s): 13.10.+q,12.10.Dm,12.15.Cc,14.60.Jj
\vfill
\noindent
--------------------------------------------\phantom{-} \hfil\break
\leftline{$^\dagger$ Talk given at the 7th Meeting of the American
Physical Society, DPF92, 10-14 November 1992,}
\leftline{\phantom{$^\dagger$}
Fermi National Accelerator Laboratory, Batavia, Illinois.}
\vglue 0.2cm
\leftline{E-mail: nardi@umiphys.bitnet}
\bigskip
\leftline{UM-TH 92--30}
                   \bigskip
\centerline{November 1992}

\bigskip \bigskip
\eject

{\elevenbf\noindent 1. Introduction}
\vglue 0.4cm
\baselineskip=16pt
\elevenrm
\noindent
The Standard Model (SM) of the electroweak interactions has achieved a
tremendous success in describing the experimental data within the
range of energies available today. The present theory also
accommodates in a satisfactory way the whole spectrum of known
particles, and only two states, the top-quark and the Higgs boson,
necessary for the consistency of the model, have not been discovered
yet. In addition a large set of experimental limits on rare processes
are explained in a satisfactory way, since the model includes the
Glashow-Iliopoulos-Maiani mechanism for suppressing flavor changing
neutral currents (FCNC) in the quark sector, and strictly forbids
lepton flavor violating (LFV) currents. All these features are quite
peculiar to the SM, and in general most of its possible extensions
predict a larger spectrum of states as well as larger rates for FCNC
processes.

Here I will analyze a new class of FCNC interactions that are
generally present in most of the extensions of the SM which predict
one (or more) additional neutral gauge boson $Z_1$ together with new
charged fermions, and that are induced by $Z_1$ interactions. In some
cases the FC couplings to the $Z_1$ are not suppressed,
and when one considers the
current limits on the corresponding FC transitions this
points toward a rather heavy  $Z_1$$^1$.
In general the
standard  $Z_0$ is expected to be
mixed with the $Z_1$. The resulting mass eigenstate will then
acquire new FC couplings proportional to the $Z_0$--$Z_1$ mixing angle$^1$.
However in this short presentation I will neglect these additional FC
effects, since due to the  tight limits implied
for the $Z_0$--$Z_1$ mixing by low energy NC and
LEP data $(\phi_{Z_0-Z_1}\lesssim 0.02\>^{2,3})$
they turn out to be less important than the effects due to
$Z_1$ exchange$^1$. In order to illustrate the power of the
constraints which can be derived from the limits on FCNC, I will apply
them to a class of E$_6$ grand unified theories (GUTs), and I will show
that once a natural range of values for the new FC parameters is
assumed, the non-observation of the decay $\mu\rightarrow eee$
tightly constraints
the mass of the $Z_1$.
\vglue 0.6cm
{\elevenbf\noindent 2. Formalism}
\vglue 0.4cm
\noindent
Once a low energy gauge group of the form
$[SU(2)_L \times U(1)_Y \times SU(3)_C]\times U_1(1)$ is assumed,
the neutral current Lagrangian in the gauge basis reads
\beq
-{\cal L}_{\rm NC}=eJ^\mu_{\rm em}A_\mu +
g_0 J_0^\mu Z_{0 \mu} + g_1 J_1^\mu Z_{1 \mu}.
\eq{2.1}
\eeq
The SM neutral gauge boson $Z_0$ couples with strength
$g_0=(4\sqrt{2} G_F M^2_{Z_0})^{1/2}$ to the usual combination of the
neutral isospin and electromagnetic currents
$
J^\mu_0=J^\mu_3-\sin^2\theta_W  J^\mu_{\rm em}.
$
Assuming that the new $U_1(1)$ originates from a GUT based on a
{\elevenit simple} group, and normalizing  the new generator $Q_1$ to the
hypercharge axis, then the $Z_1$ couples to the new $J_1$ current with
strength
$g_1 \simeq g_0 \sin\theta_W$. To ensure the absence of
anomalies for the new gauge current $J_1\,$, {\elevenit new} fermions
must be present in addition to the standard 15 {\elevenit known}
fermions per generation. Here I will assume that some of the
additional new fermions are electrically charged, and that they are
mixed with the known states. Each of the conventional {\elevenit
light} fermion mass eigenstate then corresponds to a superposition of
the known states and the new states. Conservation of the electric and color
charges forbids a mixing between gauge eigenstates with different
$U(1)_{\rm em}$ and $SU(3)_{\rm c}$ quantum numbers, implying in turn
that the corresponding currents are not modified by the presence of
the new states. In contrast the neutral isospin generator $T_3$ and
the new generator $Q_1$ are spontaneously broken, and a mixing between
states with different $t_3$ and $q_1$ eigenvalues is allowed. This
will affect the $J_3$ and $J_1$ currents$^2$ and in turn the couplings of
the light mass eigenstates to the $Z_0$ and $Z_1$.
In the gauge currents chirality is conserved too, and it is then
convenient to group the fermions with the same electric charge and
chirality $\alpha=L,R$ in a vector of the known ($\cal K$) and new
($\cal N$) gauge eigenstates$^2$  $\Psi^{o}_\alpha=(\Psi^o_{\K},
\Psi^o_{\N})_\alpha^T$. This vector is related to the corresponding vector
of the light ({\elevenit l}) and heavy ({\elevenit h}) mass eigenstates
$\Psi_\alpha=(\Psi_l,\Psi_h)_\alpha^T$ through a unitary transformation
\beq
\pmatrix{\Psi^o_{\K}\cr\Psi^o_{\N}}_\alpha = U_\alpha
\pmatrix{\Psi_l\cr\Psi_h}_\alpha \qquad{\rm where}\qquad
U_\alpha = \pmatrix{A &G\cr F & H}_\alpha ,
\qquad  \alpha=L,R.
\eq{2.6}
\eeq
In terms of the fermion mass eigenstates, the neutral current
corresponding to a (broken) generator ${\cal Q}=T_3,Q_1$ now reads
\beq
J^\mu_{\cal Q} =\sum_{\alpha=L,R}
\bar\Psi_{\alpha} \gamma^\mu U^\dagger_\alpha {\cal Q}_\alpha
U_\alpha\Psi_{\alpha},
\eq{2.8}
\eeq
where ${\cal Q}_\alpha$ represents a generic diagonal matrix of the
charges $q_\alpha=t_3(f_\alpha)$, $q_1(f_\alpha)$ for the chiral
fermion $f_\alpha$. Since we are interested in the indirect effects of
fermion mixings in the couplings of the light mass eigenstates, we
have to project $J^\mu_{\cal Q}$ onto the light components $\Psi_l$.
In the particularly simple case when the mixing is with only
one type of new fermions with the same $q_\alpha^\N$ charges,
by means of the unitarity of $U_\alpha$ we easily obtain$^2$
\beq
J^\mu_{l{\cal Q}}
               =\sum_{\alpha=L,R}
\bar \Psi_{l\alpha} \gamma^\mu \left[ q_\alpha^{\K} I +
(q_\alpha^{\N} - q_\alpha^{\K})
F^\dagger_\alpha F_\alpha \right]\Psi_{l\alpha}. \eq{2.10}
\eeq
In Eq. $(4)$
$q_\alpha^{\K}I $ represents the coupling of a particular light
fermion in the absence of mixing effects, while the second term
accounts for the modifications due to fermion mixings. The matrix
$F_\alpha^\dagger F_\alpha$ is in general not diagonal, and clearly
whenever the coefficient $(q_\alpha^{\N} - q_\alpha^{\K})$ is nonvanishing,
the off diagonal terms will induce FCNC.
We can distinguish two cases. When the mixing violates weak--isospin
$(t_3(f_\alpha^{\N}) \ne t_3(f_\alpha^{\K}))$ the $J_0$ current
is affected, and the $Z_0$ interactions will be flavor changing.
However an analysis of the fermion mass matrix$^1$ shows that the
isospin--violating fermion mixings, which  are generated via $\Delta
t_3 ={1 \over 2}$ off-diagonal mass terms, are suppressed as the ratio
of the heavy to light masses. Hence the corresponding FC terms
$(F_\alpha^\dagger F_\alpha)_{i\ne j}$ are predicted to be quite
small, and hence there is no theoretical conflict with the experimental
limits$^4$. In contrast $\Delta t_3 = 0$ mass terms induce
mixings that do not violate weak isospin,
and  it turns out that in this case no
suppression factors are present$^1$.
Since in this case $t_3(f_\alpha^{\N}) =
t_3(f_\alpha^{\K})$, clearly  the $J_0$ current is not
affected and remains flavor-diagonal. However, in general we still
have $q_1(f_\alpha^{\N}) \ne q_1(f_\alpha^{\K})$, and then the
isospin-conserving mixings can affect the $J_1$ current,
inducing sizeable FC couplings to the $Z_1$. This constrains the $Z_1$
mass to be sufficiently heavy for suppressing at
low energy the FC transitions via propagator effects.
Similarly a possible $Z_0$--$Z_1$ mixing would induce additional FC
contributions to the vertices of the $Z$ mass eigenstate, therefore
the mixing cannot be too large so as to conflict with the
experimental bounds$^1$.
\vglue 0.6cm
{\elevenbf\noindent 3. Constraints on E$_{\bf 6}$ models from
\bfmu $\rightarrow$ \bfe\bfe\bfe.}
\vglue 0.4cm
\noindent
E$_6$ GUTs are well known examples of theories where additional
fermions and new neutral gauge bosons are simultaneously present,
giving rise to the kind of effects which I have discussed. For a general
breaking of E$_6$ (rank 6) to the SM (rank 4) it is possible to define
a whole class of $Z_1$ bosons corresponding to a linear combination of the
two additional Cartan generators$^5$.
I will parametrize this combination in terms of an
angle $\beta$. Fermions are assigned to the
fundamental {\underbar {\bf 27}}
representation of the group which contains 12 additional states
for each generation, among which we have a vector doublet of new
leptons $(N\> E^-)_L^T$, $(E^+ \> N^c)_L^T$. Non-diagonal mass
terms with the standard $(\nu\> e^-)_L^T$ and $e^c_L$ leptons will
give rise respectively to $\Delta t_3=0$ and $\Delta t_3 ={1\over 2}$
mixings, and in particular will induce LFV left (L) and right (R) chiral
couplings between the first and second generation,
allowing for the decay $\mu\rightarrow eee^1$.
A very stringent experimental limit exists for this decay mode$^6$:
$Br(\mu^+\rightarrow e^+e^+e^-)<1.0\,\cdot\, 10^{-12}$.
In order to derive bounds for the $Z_1$ mass from this result I will
conservatively neglect the LFV couplings in the R sector, and assume
that the only source of the LFV interaction comes from the
$\Delta t_3 = 0$ mixing in the L sector.
I will also assume that the LFV term
${\cal F}_{e\mu}\equiv (F_L^\dagger F_L)_{e\mu}$
lies in the `natural' range
$10^{-2}$--$10^{-3}$. This assumption relies on the observation that
the CKM mixings, which are also
isospin-conserving, are numerically $> 10^{-3}$ and that the mixing
between the first and second generation is particularly large.
The limits on the $Z_1$ mass
obtained in this way$^1$ are depicted in Fig.$\, 1$ as a function of the
parameter $\beta$ that defines the particular E$_6$ boson. These
limits are indeed very strong, however,
since they depend on a specific assumption for the numerical
value of the LFV coupling, they cannot replace
the direct$^7$ bounds or other more
model independent indirect limits$^{2,3}$.
\vglue 0.5cm
\begin{figure}
\vspace{7.5cm}
\caption[]{\tenrm \baselineskip=12pt
Limits on $M_{Z_1}$ from $\mu\rightarrow eee$
for a general neutral gauge boson from E$_6$, as a
function of $\sin\beta$. The values of $\sin\beta$ corresponding to
the particularly interesting $Z_\eta$, $Z_\chi$ and $Z_\psi$ bosons,
are also shown at the top of the figure.
The limits are given for the two different values of
the LFV term ${\cal F}_{e\mu}=10^{-2}$ and $10^{-3}$.}
\label{fig1}
\end{figure}

\vglue 0.4cm
{\elevenbf\noindent 4. References \hfil}
\vglue 0.4cm \elevenrm

\end{document}